\documentclass[10pt]{acmart} 

\usepackage{array} 
\usepackage{amsmath}
\usepackage{tabularx}
\usepackage{booktabs} 
\usepackage{caption} 
\usepackage{listings}
\AtBeginDocument{%
  }

\begin{document}

\title{Malicious Code Detection in Smart Contracts via Opcode Vectorization}

\author{Huanhuan Zou}
\affiliation{
  \institution{School of Cyberspace Security, Hainan University}
  \city{Haikou}
  \country{China}
  \postcode{570228}
}

\author{Zongwei Li}
\affiliation{
  \institution{School of Cyberspace Security, Hainan University}
  \city{Haikou}
  \country{China}
  \postcode{570228}
}

\author{Xiaoqi Li}
\affiliation{
  \institution{School of Cyberspace Security, Hainan University}
  \city{Haikou}
  \country{China}
  \postcode{570228}
}
\email{csxqli@ieee.org}

\begin{abstract}
With the booming development of blockchain technology, smart contracts have been widely used in finance, supply chain, Internet of things and other fields in recent years. However, the security problems of smart contracts become increasingly prominent. Security events caused by smart contracts occur frequently, and the existence of malicious codes may lead to the loss of user assets and system crash. In this paper, a simple study is carried out on malicious code detection of intelligent contracts based on machine learning. The main research work and achievements are as follows:
Feature extraction and vectorization of smart contract are the first step to detect malicious code of smart contract by using machine learning method, and feature processing has an important impact on detection results. In this paper, an opcode vectorization method based on smart contract text is adopted. Based on considering the structural characteristics of contract opcodes, the opcodes are classified and simplified. Then, N-Gram (N=2) algorithm and TF-IDF algorithm are used to convert the simplified opcodes into vectors, and then put into the machine learning model for training. In contrast, N-Gram (N=2) algorithm and TF-IDF algorithm are directly used to quantify opcodes and put into the machine learning model training. Judging which feature extraction method is better according to the training results. Finally, the classifier chain is applied to the intelligent contract malicious code detection.
\end{abstract}





\maketitle

\section{Introduction}
With the rapid development of blockchain technology, smart contracts have seen increasingly widespread applications in finance, supply chains, insurance, and other fields. A smart contract is a computer protocol that automatically executes predefined conditions and logic. It is essentially a piece of code designed to facilitate, verify, or enforce the terms of a contract \cite{sayeed2020smart}. Although smart contracts offer advantages such as decentralization, transparency, and immutability, they still face security risks. In 2017, the Parity Multisig wallet had a fatal unauthorized access vulnerability, resulting in \$300 million being frozen \cite{metz2016biggest}. In 2018, an integer overflow vulnerability in the BEC campaign caused the instantaneous evaporation of over \$900 million. Security incidents in blockchain can be categorized into the following types: trading platform security incidents, user account and key security incidents, miner node system security incidents, smart contract security incidents, and other security incidents. Among these, smart contract security incidents account for approximately 6\%, but the losses incurred are substantial, accumulating to \$5.1 billion by the end of 2022. Therefore, addressing smart contract security issues is urgent \cite{Chen_2020_Survey, jiaoSurveyEthereumSmart2024, kumarVulnerabilitiesSmartContracts2024, weiSurveyQualityAssurance2024, zhuSurveySecurityAnalysis2024}.

Traditional static and dynamic analysis methods face certain limitations in detecting malicious code in smart contracts, including low analysis efficiency, high false-positive rates, and an inability to identify new and complex vulnerabilities \cite{Tchakounte_2022_smart, wang2024ContractsentryStaticAnalysis, grossmanPracticalVerificationSmart2024, wangContractCheckCheckingEthereum2024}. In recent years, machine learning technology has achieved remarkable results in fields such as computer vision and natural language processing, opening up new research directions for detecting malicious code in smart contracts.

Machine learning-based malicious code detection technology for smart contracts can more effectively identify and prevent vulnerabilities and malicious behaviors in smart contracts \cite{Dwivedi_2021_Legally, boi2024VulnHuntGPTSmartContract, hu2023LargeLanguageModelPowered, wei2025AdvancedSmartContract, 10.1145/3702973, sikder2025EfficientAdaptationLarge, yu2025SmartLLaMADPOReinforcedLarge}. It enables automated and intelligent auditing of smart contract code, reducing the cost of manual audits while improving audit efficiency and accuracy \cite{Ajienka_2020_empirical}. In the future, as research deepens and technology advances, machine learning-based malicious code detection technology is certain to play an even greater role in the field of smart contracts.

\section{Background}

Smart contracts are self-executing and self-verifying computer protocols designed to facilitate, verify, or enforce the negotiation of contracts between parties \cite{Barboni_2023_Smart}. Smart contracts enable trustworthy transactions without third-party intermediaries, thereby reducing the risks of fraud and default. The concept of smart contracts was first proposed by computer scientist Nick Szabo in 1994 \cite{Ghaleb_2023_AChecker, wu2025atomicity, xiao2025parallelizing, caiEnablingCompleteAtomicity2024, hanOSwapPreservingAtomicity2026, tas2024AtomicFairData}.

Smart contracts are one of the core applications of blockchain technology, implemented by deploying programmable scripts on the blockchain \cite{Sharma_2023_Mixed-Methods, aguilarSmartContractFamilies2024, wangEmpiricalAnalysisSmart2026}. These scripts can automatically trigger, execute, and verify relevant operations when specific conditions are met \cite{Tolmach_2021_Survey}. Since smart contracts run on decentralized blockchain networks, they have the following characteristics:

(1) Transparency: The code and execution results of smart contracts are visible to all participants, ensuring transaction transparency.

(2) Immutability\cite{omar2018identity}: Due to the properties of blockchain technology, once smart contracts are deployed on the blockchain, their code cannot be modified, ensuring the reliability and security of the contract.

(3) Automatic Execution: Smart contracts execute automatically when specific conditions are met, without requiring manual intervention.

(4) Decentralization: Smart contracts operate on distributed blockchain networks, independent of any centralized institution, reducing centralization risks.

\section{Method}
\subsection{Feature Extraction}
The features obtained from Ethereum smart contracts cannot be directly used as machine learning inputs, thus vectorization of smart contract features becomes a necessary step \cite{Ethereum, arceriSoundConstructionEVM2024}. This paper employs an opcode-based text vectorization method that first categorizes opcodes according to their semantic meanings in the Ethereum Virtual Machine (EVM), then transforms the simplified opcodes into vectors using N-Gram and TF-IDF algorithms.
\subsubsection{Common Feature Extraction Methods}
\begin{enumerate}
    \item \textbf{Opcodes Feature}\\
    Opcodes represent the basic execution units of smart contracts. By analyzing opcode sequences, we can capture the execution logic of contracts. Common methods include using N-Gram and TF-IDF algorithms to convert opcode sequences into vector representations.
    \item \textbf{Control Flow Feature}\\
    Control flow features describe the execution order between basic blocks in contracts. These can be extracted by constructing Control Flow Graphs (CFGs). Subsequently, graph embedding techniques (e.g., Graph2Vec) can convert CFGs into vector representations \cite{Church_2017_Word2Vec}.
    \item \textbf{Data Flow Feature}\\
    Data flow features characterize data transmission and transformation within contracts. These can be obtained by building Data Flow Graphs (DFGs), which can then be vectorized using graph embedding methods like Graph2Vec \cite{Yu_2021_Knowledge}.
    \item \textbf{Function Call Feature}\\
    Function call features represent invocation relationships between functions in contracts. This information can be extracted by analyzing smart contract source code or bytecode, then transformed into vectors using either N-Gram algorithms or graph embedding techniques \cite{li2024cobra}.
    \item \textbf{Variable and Data Type Feature}\\
    These features describe the variables and their types used in contracts. By examining source code, such features can be extracted and converted into vector representations using text representation methods like Bag-of-Words.
    \item \textbf{Code Complexity Feature}\\
    Code complexity features measure programming complexity metrics such as loop nesting depth and conditional branch counts. These can be obtained through static code analysis and directly used as numerical inputs for machine learning models.
\end{enumerate}

\subsection{Problem Analysis}
In recent years, machine learning-based smart contract vulnerability detection techniques have made significant progress. These methods analyze smart contract source code, bytecode, and opcodes to better detect potential vulnerabilities \cite{Zhuang_2020_Smart, zhuang2021smart, liASTRODetectingAccess2025, sun2025MTVHunterSmartContracts, zenggang2025NDLSCNewDeep, he2025VulnerabilityDetectionMethod}. 

Giacomo Ibba \cite{ibba2021evaluating} et al. extracted Abstract Syntax Trees (AST) from Solidity source code and then parsed the resulting AST. Based on the structural and syntactic definitions of the AST, they performed feature processing \cite{Curtis_2022_language-agnostic}. This approach leverages syntactic information from source code to extract features related to contract structure and behavior, thereby facilitating the identification of potential vulnerabilities \cite{AlDebeyan_2022_Improving}. 

Kim \cite{kim2019scanat} conducted research based on bytecode characteristics. By using encoders to highlight unique portions of raw bytecode and treating them as contract attribute tags, their method provides training data for neural network models. This bytecode-level approach can detect potential vulnerabilities that might be overlooked at the source code level.

Tann \cite{tann2018towards} performed analytical research on smart contract opcode layers. They constructed an Ethereum opcode sequence model using Long Short-Term Memory (LSTM) networks for vulnerability detection. This method utilizes opcode-level information to identify potential vulnerabilities at the underlying instruction level.

Ivica Nikolić \cite{nikolic2018finding} et al. conducted large-scale analysis of Ethereum smart contracts, revealing that source code is available for only about 1\% of approximately 1 million smart contracts. Consequently, bytecode and opcodes are substantially more accessible from Ethereum. 

\begin{figure}[t]
\centering
\includegraphics[width=0.8\textwidth]{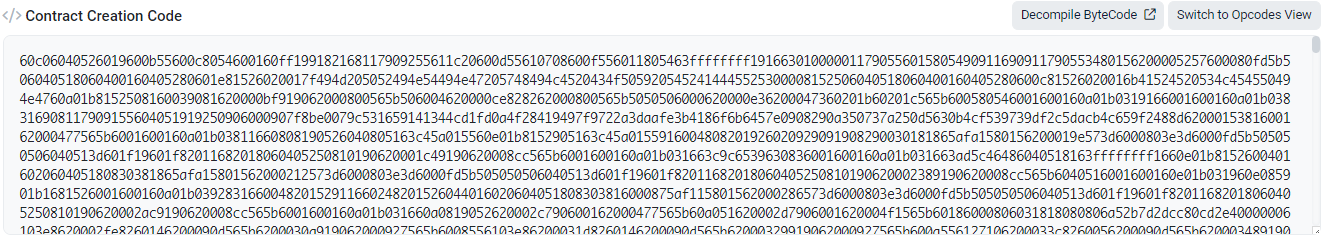}
\caption{Bytecode of Smart Contracts}
\label{fig:bytecode}
\end{figure}

\begin{figure}[h]
\centering
\includegraphics[width=0.5\textwidth]{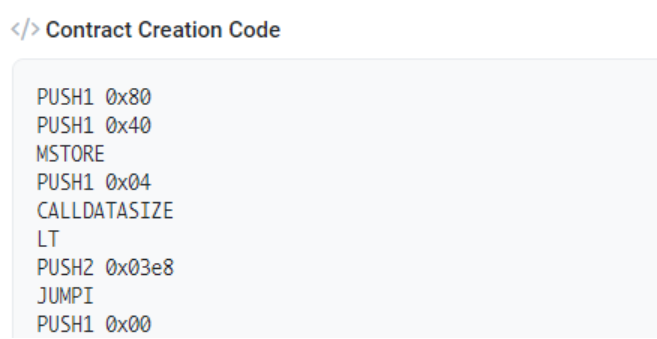}
\caption{Opcodes of Smart Contracts}
\label{fig:opcodes}
\end{figure}

As shown in Fig \ref{fig:bytecode} and \ref{fig:opcodes}, although both bytecode and opcodes are compiled from source code in Ethereum, feature extraction from bytecode may lead to semantic loss, making it difficult to adequately reflect structural features and call relationships in smart contracts. This often results in undetected vulnerabilities. Therefore, this paper extracts syntactic and semantic information characterizing vulnerabilities from smart contract opcode sequences, which better represents dataset characteristics.

For opcode vectorization, text vectorization methods can indeed serve as references. However, directly applying algorithms like Word2Vec may inadequately capture sequential relationships between opcodes, potentially compromising program context representation. Thus, this paper employs N-Gram and TF-IDF algorithms for opcode vectorization.

The N-Gram algorithm better describes static program features by capturing sequential relationships between opcodes \cite{dhaya2014detecting}. By grouping N adjacent opcodes, it preserves program context information to some extent. The TF-IDF algorithm measures the importance of specific features within the entire program, helping highlight opcodes with significant impacts on program behavior and thereby improving vector representation effectiveness.

Prior to vectorization, opcodes are categorized to prevent the curse of dimensionality and semantic loss. This simplification of dimensions during subsequent N-Gram processing and statistical feature extraction preserves contract information while reducing feature dimensionality, ultimately enhancing model training efficiency.

\subsection{Opcodes Classification} \label{sec:opcode}
Based on the semantic definitions of opcodes in the Ethereum Virtual Machine (EVM) \cite{liu2024efficient}, this paper categorizes opcodes into distinct classes. Excluding exception handling cases, opcodes with similar execution logic are grouped together. This classification approach simplifies the vectorization process while preserving critical contract information.

\subsubsection{Operational Instructions}
As Table \ref{tab:ops}, these fundamental instructions handle data processing, storage, and logging in smart contracts. They affect only stack data without altering program structure.

\begin{table}[h]
\centering
\caption{Operational Instructions}
\label{tab:ops}
\begin{tabular}{|l|p{8cm}|}
\hline
\textbf{Opcodes} & \textbf{Description} \\ \hline
PUSH1-PUSH32 & Pushes 1-32 bytes of data onto the stack \\ \hline
DUP1-DUP16 & Duplicates stack items at positions 1-16 \\ \hline
SWAP1-SWAP16 & Swaps values between stack top and positions 1-16 \\ \hline
LOG0-LOG4 & Creates log entries in smart contracts \\ \hline
\end{tabular}
\end{table}

\subsubsection{Predictable Variable Instructions}
As Table \ref{tab:predict}, used to access blockchain-specific information, storing results in memory without affecting program structure.

\begin{table}[h]
\centering
\caption{Predictable Variable Instructions}
\label{tab:predict}
\begin{tabular}{|l|p{8cm}|}
\hline
\textbf{Opcodes} & \textbf{Description} \\ \hline
BLOCKHASH & Gets the hash of specified block \\ \hline
COINBASE & Gets current block beneficiary address \\ \hline
TIMESTAMP & Gets current block timestamp \\ \hline
NUMBER & Gets current block number \\ \hline
DIFFICULTY & Gets current block difficulty \\ \hline
GASLIMIT & Gets current block gas limit \\ \hline
\end{tabular}
\end{table}

\subsubsection{Logic Instructions}
As Table \ref{tab:logic}, perform bitwise operations (AND, OR, XOR) and logical negation (NOT), impacting only the top two stack values without altering program structure.

\begin{table}[h]
\centering
\caption{Logic Instructions}
\label{tab:logic}
\begin{tabular}{|l|p{8cm}|}
\hline
\textbf{Opcodes} & \textbf{Description} \\ \hline
AND & Bitwise AND operation \\ \hline
OR & Bitwise OR operation \\ \hline
XOR & Bitwise XOR operation \\ \hline
NOT & Logical NOT operation \\ \hline
\end{tabular}
\end{table}

\subsubsection{Arithmetic Instructions}
As Table \ref{tab:arith}, execute basic mathematical operations, affecting only the top two stack values.

\begin{table}[h]
\centering
\caption{Arithmetic Instructions}
\label{tab:arith}
\begin{tabular}{|l|p{8cm}|}
\hline
\textbf{Opcodes} & \textbf{Description} \\ \hline
ADD & Addition \\ \hline
MUL & Multiplication \\ \hline
SUB & Subtraction \\ \hline
DIV & Unsigned division \\ \hline
SDIV & Signed division \\ \hline
MOD & Unsigned modulo \\ \hline
SMOD & Signed modulo \\ \hline
ADDMOD & Modular addition \\ \hline
MULMOD & Modular multiplication \\ \hline
EXP & Exponentiation \\ \hline
\end{tabular}
\end{table}

\subsubsection{Comparison Instructions}
As Table \ref{tab:comp}, perform relational operations crucial for conditional jumps. Results (0/1) are pushed onto the stack without directly affecting program structure.

\begin{table}[h]
\centering
\caption{Comparison Instructions}
\label{tab:comp}
\begin{tabular}{|l|p{8cm}|}
\hline
\textbf{Opcodes} & \textbf{Description} \\ \hline
LT & Less than (unsigned) \\ \hline
GT & Greater than (unsigned) \\ \hline
SLT & Less than (signed) \\ \hline
SGT & Greater than (signed) \\ \hline
EQ & Equality comparison \\ \hline
ISZERO & Tests if value equals zero \\ \hline
\end{tabular}
\end{table}

\subsubsection{Address and Balance Instructions}
As Table \ref{tab:addr}, provide address-related queries for transaction processing (transfers, authorizations), storing results in memory.

\begin{table}[h]
\centering
\caption{Address and Balance Instructions}
\label{tab:addr}
\begin{tabular}{|l|p{8cm}|}
\hline
\textbf{Opcodes} & \textbf{Description} \\ \hline
ADDRESS & Gets current contract address \\ \hline
BALANCE & Queries address balance \\ \hline
ORIGIN & Gets transaction origin address \\ \hline
CALLER & Gets immediate caller address \\ \hline
\end{tabular}
\end{table}

\subsubsection{CALL-family Instructions}
As Table \ref{tab:call}, enable inter-contract calls by constructing msg.data structures containing method selectors and parameters. The called contract accesses the caller address through the msg.sender global variable. Boolean return values indicate execution status (0=failure, 1=success), affecting program flow.

\begin{table}[h]
\centering
\caption{CALL-family Instructions}
\label{tab:call}
\begin{tabular}{|l|p{8cm}|}
\hline
\textbf{Opcodes} & \textbf{Description} \\ \hline
CALL & Invokes another contract's function \\ \hline
CALLCODE & Executes target contract code in current context \\ \hline
DELEGATECALL & Preserves original msg.sender and msg.value \\ \hline
STATICCALL & Performs static calls without state modification \\ \hline
\end{tabular}
\end{table}

\subsubsection{SSTORE-family Instructions}
As Table \ref{tab:storage}, handle temporary (memory) and persistent (storage) data operations. Storage modifications directly affect contract state.

\begin{table}[h]
\centering
\caption{SSTORE-family Instructions}
\label{tab:storage}
\begin{tabular}{|l|p{8cm}|}
\hline
\textbf{Opcodes} & \textbf{Description} \\ \hline
MLOAD & Loads 32-byte word from memory \\ \hline
MSTORE & Stores 32-byte word to memory \\ \hline
MSTORE8 & Stores single byte to memory \\ \hline
SLOAD & Loads word from storage \\ \hline
SSTORE & Stores word to storage \\ \hline
\end{tabular}
\end{table}

\subsubsection{Termination Instructions}
As Table \ref{tab:term}. handle execution conclusions including normal termination, exceptions, and self-destruction. Subsequent execution requires JUMP instructions, directly affecting program structure.

\begin{table}[h]
\centering
\caption{Termination Instructions}
\label{tab:term}
\begin{tabular}{|l|p{8cm}|}
\hline
\textbf{Opcodes} & \textbf{Description} \\ \hline
RETURN & Terminates with specified return data \\ \hline
STOP & Halts contract execution \\ \hline
REVERT & Reverts state changes with return data \\ \hline
INVALID & Triggers on undefined opcodes \\ \hline
SELFDESTRUCT & Destroys current contract \\ \hline
\end{tabular}
\end{table}

\subsubsection{Jump Instructions}
As Table \ref{tab:jump}, implement control flow logic through conditional/unconditional jumps, directly modifying program execution paths.

\begin{table}[h]
  \centering
  \caption{Jump Instructions}
  \label{tab:jump}
  \begin{tabular}{|l|p{8cm}|}
  \hline
  \textbf{Opcodes} & \textbf{Description} \\ \hline
  JUMPDEST & Marks valid jump destination \\ \hline
  JUMP & Unconditional jump \\ \hline
  JUMPI & Conditional jump (executes if top stack value $\neq$ 0) \\ \hline
  \end{tabular}
  \end{table}

\subsection{Opcode Vectorization}
In the analysis of smart contract opcodes, drawing an analogy to word vectorization techniques in natural language processing is an effective approach. Mapping opcodes into vector space can reveal relationships and similarities between opcodes, thereby providing valuable information for subsequent malicious code detection and vulnerability identification tasks. Using the N-Gram algorithm can preserve some sequential information of opcodes and capture associations between adjacent opcodes. This is because the N-Gram algorithm divides text into sequences of N consecutive elements to capture collocations and relationships between these elements. Thus, we can obtain a feature representation that reflects the contextual relationships between opcodes. Additionally, the TF-IDF algorithm can be used to measure the importance of opcodes within the entire program. By calculating the term frequency (TF) and inverse document frequency (IDF) of each opcode, we can assign a weight to each opcode to highlight those that have a greater impact on the program's behavior.

\subsubsection{N-Gram Algorithm}
The N-Gram algorithm is a statistical language model-based method used to analyze and represent relationships between adjacent elements (such as words or characters) in text. The core idea of the N-Gram algorithm is to divide text into sequences of N consecutive elements to capture collocations and relationships between these elements. Each sequence is called a gram, and each gram represents a dimension of the feature vector. Considering the large number of smart contract opcodes, we set N=2 (i.e., bigram) to keep the dimensionality manageable. The specific steps are as follows:

\begin{enumerate}
    \item First, split the smart contract's opcode sequence into pairs of adjacent opcodes (i.e., bigrams). For example, given the opcode sequence A B C D, the generated bigrams are: (A,B), (B,C), (C,D).
    \item Calculate the frequency of each bigram in the opcode sequence. This can be achieved by traversing the entire opcode sequence and counting the occurrences of each bigram.
    \item Convert the bigram frequency data into vector representation. Create a vector where each element corresponds to a possible bigram, and set the value of each element to the frequency of the corresponding bigram in the opcode sequence. For example, if all possible bigrams are \{(A,B), (A,C), (B,C), (B,D), (C,D)\}, the vector representation of the opcode sequence A B C D might be [1, 0, 1, 0, 1].
\end{enumerate}

\subsubsection{TF-IDF Algorithm}
The TF-IDF algorithm is a common text mining method used to measure the importance of words in a document. It is based on two main concepts: Term Frequency (TF) and Inverse Document Frequency (IDF). By combining these two concepts, the TF-IDF algorithm can highlight words that appear frequently in a specific document but are rare across the entire document set, thereby assigning them higher weights.

Term Frequency (TF) refers to the number of times a term appears in a document. Generally, the higher the term frequency, the greater its importance in the document. The formula is as follows:

\begin{equation}
 TF(t, D) = \frac{ t }{ D} 
\end{equation}

where \( t \) represents a specific bigram in a smart contract, and \( D \) represents the total number of bigrams in that smart contract.

Inverse Document Frequency (IDF) is a correction applied to term frequency to reduce the influence of high-frequency terms (here, high-frequency bigrams). The formula is as follows:
\begin{equation}
 IDF(t, N) = \log\left(\frac{N}{n_t + 1}\right) 
\end{equation}
where \( N \) is the total number of smart contracts in the training set, \( n_t \) is the number of smart contracts containing a specific bigram, and \( n_t + 1 \) is added to prevent division by zero in case no smart contract contains the bigram.

The TF-IDF value is calculated as:
\begin{equation}
 TF\text{-}IDF(t, D, N) = TF(t, D) \times IDF(t, N) 
\end{equation}

\subsubsection{Opcodes Simplification}
To reduce dimensionality, we simplify the number of opcodes before applying the N-Gram algorithm. Based on the roles and impacts of various opcodes discussed in Section \ref{sec:opcode}, this paper adopts the following simplification rules (Table \ref{tab:simplification}):

After simplification, 35 opcodes remain: \texttt{PUSH}, \texttt{DUP}, \texttt{SWAP}, \texttt{LOG}, \texttt{COUNT}, \texttt{PREDICT}, \texttt{JUDGE}, \texttt{COMPARISON}, \texttt{ADDRESS}, \texttt{JUMPDEST}, \texttt{JUMP}, \texttt{JUMPI}, \texttt{CALLVALUE}, \texttt{ALLDATALOAD}, \texttt{CALLDATASIZE}, \texttt{CALLDATACOPY}, \texttt{CALL}, \texttt{CODESIZE}, \texttt{CODECOPY}, \texttt{GASPRICE}, \texttt{CREATE}, \texttt{EXTCODESIZE}, \texttt{EXTCODECOPY}, \texttt{GAS}, \texttt{STOP}, \texttt{EQ}, \texttt{CALLCODE}, \texttt{DELEGATECALL}, \texttt{SELEGATECALL}, \texttt{SELFDESTRUCT}, \texttt{REVERT}, \texttt{MSTORE}, \texttt{SHA3}, \texttt{POP}, \texttt{RETURN}.

\begin{table}[h]
\centering
\caption{Instruction Simplification Mapping}
\label{tab:simplification}
\begin{tabular}{|c|c|}
\hline
\textbf{Original Instruction} & \textbf{Simplified Instruction} \\ \hline
Operation Instructions & PUSH1-PUSH16 → PUSH \\
                       & DUP1-DUP16 → DUP \\
                       & SWAP1-SWAP16 → SWAP \\
                       & LOG0-LOG4 → LOG \\ \hline
Arithmetic Instructions & All → COUNT \\ \hline
Predictable Instructions & All → PREDICT \\ \hline
Judgment Instructions & All → JUDGE \\ \hline
Comparison Instructions & All → COMPARISON \\ \hline
Address/Wallet Instructions & All → ADDRESS \\ \hline
Instructions affecting program structure & No simplification \\ \hline
Instructions not closely related to vulnerabilities & Removed \\ \hline
\end{tabular}
\end{table}

\begin{figure}[h]
\centering
\includegraphics[width=0.8\textwidth]{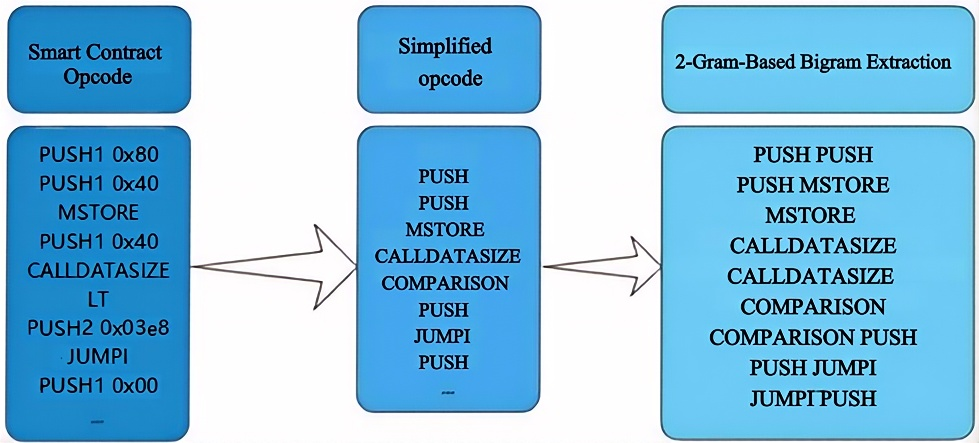}
\caption{Flowchart of bigram extraction from simplified opcodes}
\label{fig:bigram_extraction}
\end{figure}

The 2-Gram algorithm is used to extract bigrams from the simplified contract opcodes. The extraction process is shown in Fig \ref{fig:bigram_extraction}.

This process ultimately extracts 1225-dimensional bigrams, which effectively reflects the relationships between opcodes while avoiding dimensionality issues during model training.

Since smart contracts vary in length, the types of extracted bigrams may differ. If we only calculate TF-IDF values for bigrams present in a particular smart contract's opcodes, the resulting vector dimensions would vary across contracts, creating difficulties for subsequent model training. To address this, we standardize the vector dimension to 1225 for all smart contracts. Each dimension corresponds to a specific bigram, and if a particular bigram doesn't appear in a contract's opcodes, its corresponding TF-IDF value is set to 0.

\subsubsection{Pseudocode Implementation}
Table \ref{tab:bigram_construction} extracts all possible bigrams from the simplified opcodes, ultimately generating 1225-dimensional bigrams.

\begin{table}[h]
\centering
\caption{Pseudocode for Constructing Bigrams}
\label{tab:bigram_construction}
\begin{tabular}{|l|}
\hline
\textbf{Construct all possible bigram combinations} \\
\hline
1: Define opcode list opcodes = ['PUSH','DUP','SWAP',...] \\
2: Define an empty list all\_bigrams for all possible combinations \\
3: \textbf{for} each element a in opcodes: \\
4: \quad \textbf{for} each element b in opcodes: \\
5: \quad \quad Concatenate opcode a and opcode b with a space \\
6: \quad \quad Add the concatenated string to all\_bigrams list \\ 
\hline
\end{tabular}
\end{table}

Table \ref{tab:opcode_classification} categorizes smart contract opcodes according to the classification rules.

\begin{table}[h]
\centering
\caption{Pseudocode for Opcode Classification Algorithm}
\label{tab:opcode_classification}
\begin{tabular}{|l|}
\hline
\textbf{Algorithm 1: Opcode Classification} \\
\textbf{Input:} Pre-simplified opcode \\
\textbf{Output:} Classified opcode \\
\hline
Function simplify\_opcode(opcode): \\
\quad \textbf{if} opcode starts with "PUSH": \textbf{return} "PUSH" \\
\quad \textbf{elif} opcode starts with "DUP": \textbf{return} "DUP" \\
\quad \textbf{elif} opcode starts with "SWAP": \textbf{return} "SWAP" \\
\quad \textbf{elif} opcode starts with "LOG": \textbf{return} "LOG" \\
\quad \textbf{elif} opcode belongs to arithmetic instructions: \textbf{return} "COUNT" \\
\quad \textbf{elif} opcode belongs to predictable instructions: \textbf{return} "PREDICT" \\
\quad \textbf{elif} opcode belongs to judgment instructions: \textbf{return} "JUDGE" \\
\quad \textbf{elif} opcode belongs to comparison instructions: \textbf{return} "COMPARISON" \\
\quad \textbf{elif} opcode belongs to address/wallet instructions: \textbf{return} "ADDRESS" \\
\quad \textbf{else: return} opcode \\
\hline
\end{tabular}
\end{table}

Table \ref{tab:ngram_algorithm} divides each smart contract's opcodes using the 2-Gram algorithm.

\begin{table}[h]
\centering
\caption{Pseudocode for N-Gram Algorithm}
\label{tab:ngram_algorithm}
\begin{tabular}{|l|}
\hline
\textbf{Algorithm 2: N-Gram Algorithm for Smart Contract Opcodes} \\
\textbf{Input:} Smart contract opcodes \\
\textbf{Output:} Bigram frequency list \\
\hline
Function bigrams\_frequency(contracts): \\
\quad Initialize an empty bigram frequency list bigrams\_freq \\
\quad \textbf{for} each contract in contracts: \\
\quad \quad Apply simplify\_opcode function to each opcode in contract \\
\quad \quad Create bigrams contract\_bigrams from simplified opcodes \\
\quad \quad Add contract\_bigrams to bigrams\_freq list \\
\quad \textbf{return} bigrams\_freq \\
\hline
\end{tabular}
\end{table}

Table \ref{tab:tfidf_algorithm} calculates the TF-IDF values for each smart contract's bigrams.

\begin{table}[h]
\centering
\caption{Pseudocode for TF-IDF Algorithm}
\label{tab:tfidf_algorithm}
\begin{tabular}{|l|}
\hline
\textbf{Algorithm 3 TF-IDF Algorithm} \\
\textbf{Input:} Smart contract opcodes \\
\textbf{Output:} TF-IDF matrix \\
\hline
Function TF\_IDF(contracts): \\
\quad Call bigrams\_frequency to compute bigram frequencies \\
\quad Create index mapping bigram\_to\_index for all bigrams \\
\quad Initialize TF-IDF vectorizer with vocabulary=bigram\_to\_index, lowercase=False \\
\quad Fit and transform bigrams\_freq using the vectorizer \\
\quad \textbf{return} TF-IDF matrix \\
\hline
\end{tabular}
\end{table}

\section{Experiment}
\subsection{Dataset Collection}
Etherscan \cite{etherscan} is an Ethereum blockchain explorer and analytics platform. It allows users to query and retrieve information about transactions, addresses, blocks, tokens (including ERC-20 and ERC-721 tokens), and smart contracts on the Ethereum network. Etherscan provides developers, researchers, and other participants in the Ethereum ecosystem with a user-friendly interface to track and analyze activities on the Ethereum blockchain.

For this experiment, we manually downloaded the source code of 500 smart contracts published on \url{https://etherscan.io/contractsVerified}. We then decoded the bytecode into opcodes using an open-source opcode tool, copied each smart contract's opcodes, and saved them as corresponding ".txt" files.

\subsection{Label Annotation}
\begin{figure}[h]
\centering
\includegraphics[width=0.8\textwidth]{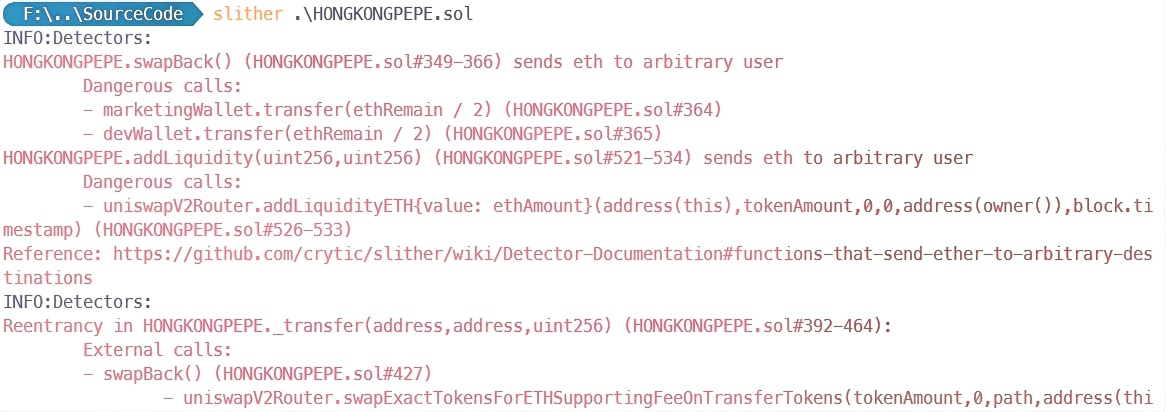}
\caption{Vulnerability detection in smart contracts using Slither}
\label{fig:slither}
\end{figure}

We used the Ethereum static analysis tool Slither \cite{feist2019slither} to detect vulnerabilities in smart contracts. After configuring the Slither tool and selecting the compiler version (solc) according to the pragma solidity \^{}0.6.0 statement in the smart contract code, we performed the detection. The detection results for one sample smart contract (as shown in Figure \ref{fig:slither}) revealed both access control vulnerabilities and reentrancy attack vulnerabilities.

Among the 500 smart contracts analyzed, 80 were found to contain vulnerabilities.

\begin{table}[h]
\centering
\caption{Vulnerability distribution in the dataset}
\label{tab:vulnerability_dist}
\begin{tabular}{|l|c|}
\hline
\textbf{Vulnerability Type} & \textbf{Number of Contracts} \\ \hline
Access Control Vulnerabilities & 37 \\ \hline
Reentrancy Attack Vulnerabilities & 58 \\ \hline
\end{tabular}
\end{table}

As Table \ref{tab:vulnerability_dist}We divided the dataset into 70\% for training and 30\% for testing. Based on Slither's detection results, for each smart contract, we labeled it as 1 if a particular vulnerability existed, and 0 otherwise.

For each vulnerability type, we calculate detection performance metrics derived from confusion matrix analysis. The overall detection effectiveness is subsequently determined by averaging these metrics across all vulnerability categories. The evaluation system employs four key indicators: average accuracy (acc), average precision (pre), average recall (rec), and average F1-score (F1), formally defined as follows:

\begin{equation}
\text{Accuracy} = \frac{TP + TN}{TP + TN + FP + FN}
\end{equation}
where Accuracy measures the proportion of correct predictions among all predictions made, reflecting the model's overall detection capability.

\begin{equation}
\text{Precision} = \frac{TP}{TP + FP}
\end{equation}
Precision quantifies the ratio of correctly identified vulnerabilities to all positive predictions, indicating the model's prediction reliability.

\begin{equation}
\text{Recall} = \frac{TP}{TP + FN}
\end{equation}
Recall represents the fraction of actual vulnerabilities successfully detected, measuring the model's capability to identify all relevant cases.

\begin{equation}
\text{F1-score} = \frac{2 \times \text{Precision} \times \text{Recall}}{\text{Precision} + \text{Recall}}
\end{equation}
The F1-score provides a harmonic mean of precision and recall, offering a balanced assessment of model performance.

In these formulations:
\textbf{TP (True Positive)}: Contracts containing specific vulnerabilities that are correctly identified. 
\textbf{TN (True Negative)}: Vulnerability-free contracts properly recognized as secure. 
\textbf{FP (False Positive)}: Secure contracts erroneously flagged as vulnerable.
\textbf{FN (False Negative)}: Vulnerable contracts that remain undetected by the system.

\subsection{Experimental Results and Analysis}
In this experiment, we employed two feature extraction methods:

\begin{itemize}
    \item \textbf{Method 1}: Vectorization based on smart contract opcode text.
    \item \textbf{Method 2}: Using 2-Gram algorithm to split opcode sequences into bigram features, then calculating TF-IDF values for each dimension as vector input.
\end{itemize}

We trained five models using these methods: SVM, Decision Tree, Random Forest, KNN, and Logistic Regression.

Due to the limited size of the smart contract opcode dataset, the access control vulnerability could not produce results in any of the five models. While reentrancy attack vulnerabilities yielded results, the detection performance across models was not significantly distinguishable, with substantial observed variance.

As shown in Table \ref{tab:method1_results}, the small dataset size led to identical training results across all five models when using Method 1.

\begin{table}[h]
\centering
\caption{Training results of different models using Method 1}
\label{tab:method1_results}
\begin{tabular}{|l|c|c|c|c|}
\hline
\textbf{Machine Learning Model} & \textbf{Accuracy} & \textbf{Precision} & \textbf{Recall} & \textbf{F1-score} \\ \hline
Support Vector Machine & 0.583 & 0.583 & 1.0 & 0.737 \\ \hline
Decision Tree & 0.583 & 0.583 & 1.0 & 0.737 \\ \hline
Random Forest & 0.583 & 0.583 & 1.0 & 0.737 \\ \hline
K-Nearest Neighbors & 0.583 & 0.583 & 1.0 & 0.737 \\ \hline
Logistic Regression & 0.583 & 0.583 & 1.0 & 0.737 \\ \hline
\end{tabular}
\end{table}

Table \ref{tab:method2_results} presents the results using Method 2, where the Decision Tree model demonstrates comparatively better performance.

\begin{table}[h]
\centering
\caption{Training results of different models using Method 2}
\label{tab:method2_results}
\begin{tabular}{|l|c|c|c|c|}
\hline
\textbf{Machine Learning Model} & \textbf{Accuracy} & \textbf{Precision} & \textbf{Recall} & \textbf{F1-score} \\ \hline
Support Vector Machine & 0.583 & 0.583 & 1.0 & 0.737 \\ \hline
Decision Tree & 0.667 & 0.65 & 0.929 & 0.765 \\ \hline
Random Forest & 0.542 & 0.565 & 0.929 & 0.703 \\ \hline
K-Nearest Neighbors & 0.583 & 0.583 & 1.0 & 0.737 \\ \hline
Logistic Regression & 0.583 & 0.583 & 1.0 & 0.737 \\ \hline
\end{tabular}
\end{table}

\begin{figure}[h]
\centering
\includegraphics[width=0.5\textwidth]{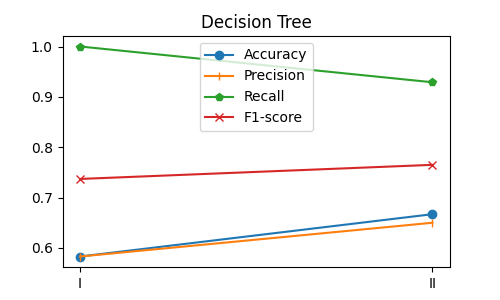}
\caption{Comparison of model training results between two methods}
\label{fig:training_results}
\end{figure}

Figures \ref{fig:training_results} visually compares the performance of Decision Tree and Random Forest models under both feature extraction methods.

Key observations from the results:
\begin{itemize}
    \item Most models showed identical performance with both methods.
    \item Decision Tree using Method 2 achieved better Accuracy and F1-score than Method 1.
    \item Random Forest with Method 1 demonstrated superior performance across all metrics compared to Method 2.
\end{itemize}

The imbalanced dataset significantly impacts training outcomes, leading to substantial variance in results. Method 2's bigram-based feature extraction showed particular promise with Decision Tree models, suggesting this approach may better capture relevant patterns when sufficient training data is available.

\section{Classifier Chain-Based Malicious Code Detection in Smart Contracts}

In practical applications, many scenarios require handling multi-label problems, such as detecting multiple types of malicious code in smart contracts. Decomposing multi-label problems into multiple independent binary classification problems may indeed ignore potential relationships between labels, thereby reducing the classifier's detection capability. The classifier chain model is an effective method for solving multi-label learning problems, which fully considers the correlations between labels while maintaining acceptable computational complexity. By linking binary classifiers in a directed structure, the classifier chain makes individual label predictions serve as features for other classifiers. This approach can capture dependencies between labels and improve the classifier's detection capability. Applying the classifier chain model to smart contract malicious code detection is a reasonable choice. In practice, we collect a multi-labeled smart contract dataset and then train it using the classifier chain model. The trained model will be able to simultaneously identify multiple types of vulnerabilities in a single contract, thereby improving detection accuracy and efficiency.
The overall process of malicious code detection is shown in the following figure \ref{fig:cc}:
\begin{figure}[h]
    \centering
    \includegraphics[width=0.4\textwidth]{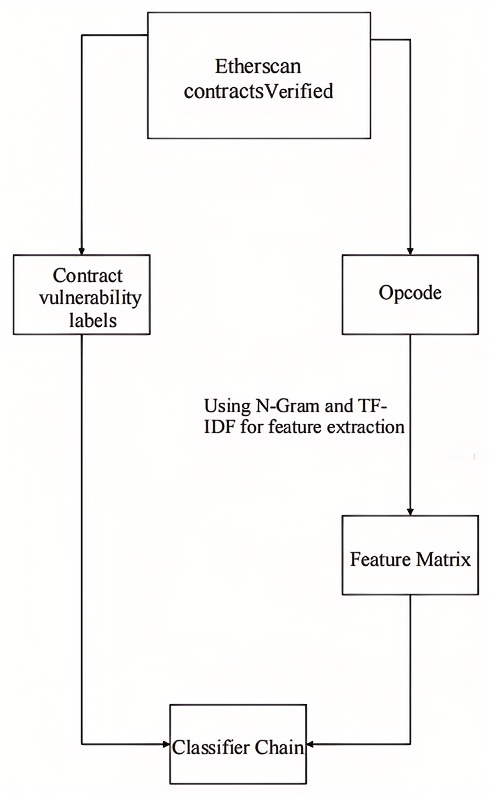}
    \caption{Overall flowchart of malicious code detection}
    \label{fig:cc}
\end{figure}
\subsection{Principles of Classifier Chains}
The classifier chain is a method for handling multi-label learning problems. Multi-label learning refers to cases where each sample in a dataset may have multiple related labels. The goal of classifier chains is to capture correlations between labels to improve the predictive performance of classifiers. In multi-label learning, classifier chains connect multiple binary classifiers in a directed structure, making individual label predictions serve as features for other classifiers. The working principle of classifier chains is as follows:
\begin{enumerate}
    \item For a given multi-label dataset, first determine an order of labels, such as sorting them by frequency of occurrence.
    \item For each label, train an independent binary classifier that predicts whether the current label exists. For the first label, the binary classifier uses only the original features as input. For subsequent labels, the binary classifier uses both the original features and the predictions of all preceding labels as input.
    \item For predicting a new sample, first use the first binary classifier to predict the first label. Then, combine the prediction result with the original features and pass them as input to the second binary classifier. This process continues until all labels are predicted.
\end{enumerate}

\subsection{Experimental Results and Analysis}
This paper employs the "ClassifierChain" class with Random Forest classifiers (RandomForestClassifier) as base classifiers. 
When comparing with other supervised algorithm models, we continue to use the confusion matrix. Since the classifier chain handles multi-label learning problems, the vulnerability labels here include reentrancy attacks and access control vulnerabilities. The feature vectors are processed using a smart contract opcode vectorization method based on contract text. Four metrics—Accuracy, Precision, Recall, and F1-Score—are used to evaluate the effectiveness of smart contract malicious code detection.

The training results of the classifier chain are shown in Table~\ref{tab:results}:

\begin{table}[h]
\centering
\caption{Training Results}
\label{tab:results}
\begin{tabular}{|c|c|c|c|c|}
\hline
Model & Accuracy & Precision & Recall & F1-Score \\
\hline
Classifier Chain & 0.333 & 0.272 & 0.467 & 0.343 \\
\hline
\end{tabular}
\end{table}

From the training results, we can observe that the performance of the classifier chain model is suboptimal. The accuracy, precision, recall, and F1-Score are all relatively low. This outcome is likely attributed to imbalanced dataset collection.

\section{Related Work}
In machine learning-based malicious code detection, feature extraction and representation are crucial \cite{liu2025sok}. Researchers attempt to extract various features from smart contract code, including syntactic features, semantic features, control flow, and data flow \cite{bu2025smartbugbert}. These features can be represented using techniques such as the bag-of-words model, word vector embeddings, and graph representation learning \cite{Asudani_2023_Impact}. Recent studies further demonstrate the importance of transaction pattern analysis, as seen in \cite{li2024detecting}, which detects malicious Web3 accounts through transaction graph analysis. Effective feature extraction and representation methods help improve the detection performance of machine learning models \cite{li2024guardians}.

In a 2019 study by Wang et al. \cite{momeni2019machine}, ASTs were extracted from Solidity source code as an intermediate representation of programs for smart contract vulnerability detection. Subsequent work by \cite{mao2024scla} introduced LLM-based approaches for automated smart contract summarization, enhancing AST utilization through control flow prompts. They experimented with four common binary classifiers, including Support Vector Machines (SVM), Neural Networks (NNs), Random Forests (RF), and Decision Trees (DT), to train models. In this study, they built and evaluated a total of 184 machine learning models to predict 16 types of vulnerabilities. By comparing the performance of these models, researchers could identify the most suitable methods for smart contract vulnerability detection \cite{bu2025enhancing}. Such research is highly significant for improving the security and reliability of smart contracts \cite{Wang_2022_Unified}.

In 2021, Xu et al. \cite{xu2021novel} extracted corresponding abstract syntax trees from the source code of each vulnerability dataset, as well as from a large dataset. They then extracted feature vectors based on subnodes shared by the two ASTs. Finally, they used the K-Nearest Neighbor (KNN) algorithm and Stochastic Gradient Descent (SGD) to train models. Both methods leveraged abstract syntax trees as the basis for feature extraction, avoiding the need to set up complex patterns or execute them on Ethereum. Recent advancements like \cite{li2024stateguard} and \cite{li2024defitail} further extend this paradigm by detecting state defects in decentralized exchanges and analyzing cross-contract execution patterns. This makes these methods relatively simple for professionals to use. However, a limitation of these approaches is that obtaining source code is more difficult compared to other contract attributes, reducing their practical utility for third-party detection. Alternative solutions like \cite{li2020characterizing} and \cite{10707457} address this challenge by characterizing account behaviors and detecting malicious activities through gas usage patterns.

The classifier chain method has become a popular approach for solving multi-label learning problems. By linking binary classifiers in a directed structure, individual label predictions serve as features for other classifiers \cite{li2021hybrid}. This method is flexible and effective, achieving state-of-the-art experimental performance across multiple datasets and multi-label evaluation metrics \cite{li2017discovering}. In recent years, many studies have explored the theoretical foundations of classifier chains and continued to improve them.

The Bayesian classifier chain algorithm proposed by Zaragoza et al. \cite{zaragoza2011bayesian} combines Bayesian networks with classifier chains. By inferring a Tree Augmented Network (TAN) from the training dataset, the correlations between labels are represented in a tree structure. Then, a node is randomly selected from the TAN as the root, and different paths from this root are used to form different chains.

Hernandez et al. \cite{hernandez2012hybrid} proposed a hybrid binary-chain multi-label classifier that defines different chains based on the correlation coefficients calculated between each pair of labels. Each chain consists of labels identified as strongly positively correlated. Kumar et al. \cite{kumar2013beam} adopted a kernel-target alignment technique with tunable input parameter beam width to determine suitable chain orders. By performing beam search on a tree, where each distinct path represents a different label permutation, they addressed the label ordering problem.

Research on machine learning-based malicious code detection in smart contracts has made progress but still faces many challenges and problems. The integration of AI and blockchain for privacy preservation \cite{li2023overview}, the detection of emerging threats like NFT wash trading \cite{niu2024unveiling}, and the need to characterize complex ecosystems like Solana NFTs \cite{kong2024characterizing} highlight the evolving nature of this field. Additionally, emerging tools like \cite{wang2024smart} statistically quantify external data dependencies in real-world contracts, while \cite{li2025scalm} demonstrates how LLMs can automate bad practice detection, pointing to future research directions.

\section{Conclusion and Future Work}
\subsection{Conclusion}
Smart contracts containing malicious code may lead to significant security issues and asset losses, necessitating effective detection and prevention mechanisms. Consequently, machine learning-based malicious code detection techniques for smart contracts have garnered widespread application. The main contributions of this work include:

\begin{enumerate}
    \item This paper provides a brief introduction to machine learning-based malicious code detection in smart contracts, with a focus on an opcode vectorization method based on contract text.
    \item It discusses five machine learning models and a classifier chain-based approach for malicious code detection.
\end{enumerate}

The experimental data reveals significant issues with the collected dataset, leading to substantial errors in training results. However, in the realm of smart contracts, malicious contracts are relatively rare and difficult to collect. This results in severe dataset imbalance, with an abundance of normal contracts and very few malicious ones. Such imbalance may introduce bias during model training, thereby affecting detection performance. Additionally, dataset quality is another critical issue, including label errors and duplicate samples, which may further degrade model performance.

\subsection{Future Work}
This study utilizes an opcode-based feature vectorization method, but due to the very limited smart contract opcode dataset available, the effectiveness of this feature extraction method remains unverified. Future work will focus on constructing a large and representative dataset of smart contract samples.

\begin{enumerate}
    \item Developing web crawlers to extract deployed contract code from Ethereum Etherscan could save considerable time. Current smart contract vulnerability detection tools cannot perform cross-version detection. For example, if a contract is written in Solidity versions 0.6.0 and 0.7.0, it becomes unsuitable as a sample, leading to significant time waste during manual screening.
    \item To address the lack of malicious contract samples in existing datasets, semi-supervised and unsupervised learning methods should be explored to leverage large quantities of unlabeled contract samples for training.
    \item Privacy protection and security challenges must be addressed to ensure that machine learning-based malicious code detection systems do not leak sensitive information or become vulnerable to adversarial attacks.
\end{enumerate}







\bibliographystyle{ACM-Reference-Format}
\bibliography{ref}

@inproceedings{momeni2019machine,
  title     = {Machine learning model for smart contracts security analysis},
  author    = {Momeni, Pouyan and Wang, Yu and Samavi, Reza},
  booktitle = {Proceedings of the 17th international conference on privacy, security and trust (PST)},
  pages     = {1--6},
  year      = {2019}
}

@article{xu2021novel,
  title   = {A novel machine learning-based analysis model for smart contract vulnerability},
  author  = {Xu, Yingjie and Hu, Gengran and You, Lin and Cao, Chengtang},
  journal = {Security and Communication Networks},
  volume  = {2021},
  number  = {1},
  pages   = {5798033},
  year    = {2021}
}

@misc{zaragoza2011bayesian,
  title  = {Bayesian chain classifiers for multidimensional classification},
  author = {Zaragoza, Julio H and Sucar, Luis Enrique and Morales, Eduardo F and Larra{\~n}aga M{\'u}gica, Pedro Mar{\'\i}a and Bielza Lozoya, Mar{\'\i}a Concepci{\'o}n},
  year   = {2011}
}

@inproceedings{hernandez2012hybrid,
  title     = {Hybrid binary-chain multi-label classifiers},
  author    = {Hernandez-Leal, Pablo and Orihuela-Espina, Felipe and Sucar, Enrique and Morales, Eduardo F},
  booktitle = {Proceedings of the 6th European Workshop Probabilistic Graphical Models},
  year      = {2012}
}

@article{kumar2013beam,
  title   = {Beam search algorithms for multilabel learning},
  author  = {Kumar, Abhishek and Vembu, Shankar and Menon, Aditya Krishna and Elkan, Charles},
  journal = {Machine learning},
  volume  = {92},
  pages   = {65--89},
  year    = {2013}
}

@article{sayeed2020smart,
  title   = {Smart contract: Attacks and protections},
  author  = {Sayeed, Sarwar and Marco-Gisbert, Hector and Caira, Tom},
  journal = {Ieee Access},
  volume  = {8},
  pages   = {24416--24427},
  year    = {2020}
}

@misc{metz2016biggest,
  title     = {The biggest crowdfunding project ever-the DAO-is kind of a mess},
  author    = {Metz, Cade},
  year      = {2016},
  publisher = {Wired}
}

@inproceedings{omar2018identity,
  title     = {Identity management in IoT networks using blockchain and smart contracts},
  author    = {Omar, Ahmad Sghaier and Basir, Otman},
  booktitle = {2018 IEEE International Conference on Internet of Things (iThings) and IEEE Green Computing and Communications (GreenCom) and IEEE Cyber, Physical and Social Computing (CPSCom) and IEEE Smart Data (SmartData)},
  pages     = {994--1000},
  year      = {2018}
}

@inproceedings{Curtis_2022_language-agnostic,
  title     = {On Language-Agnostic Abstract-Syntax Trees: Student Research Abstract},
  booktitle = {Proceedings of the  {{37th ACM}}/{{SIGAPP Symposium}} on {{Applied Computing}} (SAC)},
  author    = {Curtis, Jacob},
  year      = {2022},
  pages     = {1619--1625}
}

@inproceedings{Zhuang_2020_Smart,
  title     = {Smart {{Contract Vulnerability Detection}} Using {{Graph Neural Network}}},
  booktitle = {Proceedings of the {{29th International Joint Conference}} on {{Artificial Intelligence}} (IJCAI)},
  author    = {Zhuang, Yuan and Liu, Zhenguang and Qian, Peng and Liu, Qi and Wang, Xiang and He, Qinming},
  year      = {2020},
  pages     = {3283--3290}
}

@article{Ethereum,
  title   = {Ethereum: A secure decentralised generalised transaction ledger},
  author  = {Wood, Gavin and others},
  journal = {Ethereum project yellow paper},
  volume  = {151},
  number  = {2014},
  pages   = {1--32},
  year    = {2014}
}

@article{Chen_2020_Survey,
  title   = {A {{Survey}} on {{Ethereum Systems Security}}: {{Vulnerabilities}}, {{Attacks}}, and {{Defenses}}},
  author  = {Chen, Huashan and Pendleton, Marcus and Njilla, Laurent and Xu, Shouhuai},
  year    = {2020},
  journal = {ACM Computing Surveys},
  volume  = {53},
  number  = {3},
  pages   = {1--43}
}

@inproceedings{Sharma_2023_Mixed-Methods,
  author    = {Tanusree Sharma and Zhixuan Zhou and Andrew Miller and Yang Wang},
  title     = {A {Mixed-Methods} Study of Security Practices of Smart Contract Developers},
  booktitle = {Proceedings of the {32nd USENIX Security Symposium} (USENIX Security)},
  year      = {2023},
  pages     = {2545--2562}
}

@inproceedings{Barboni_2023_Smart,
  title     = {Smart Contract Testing: Challenges and Opportunities},
  booktitle = {Proceedings of the {{5th International Workshop}} on {{Emerging Trends}} in {{Software Engineering}} for {{Blockchain}} (WETSEB)},
  author    = {Barboni, Morena and Morichetta, Andrea and Polini, Andrea},
  year      = {2023},
  pages     = {21--24}
}

@article{Tolmach_2021_Survey,
  title   = {A {{Survey}} of {{Smart Contract Formal Specification}} and {{Verification}}},
  author  = {Tolmach, Palina and Li, Yi and Lin, Shang-Wei and Liu, Yang and Li, Zengxiang},
  year    = {2021},
  journal = {ACM Computing Surveys},
  volume  = {54},
  number  = {7},
  pages   = {1--38}
}

@article{Dwivedi_2021_Legally,
  title   = {Legally {{Enforceable Smart-Contract Languages}}: {{A Systematic Literature Review}}},
  author  = {Dwivedi, Vimal and Pattanaik, Vishwajeet and Deval, Vipin and Dixit, Abhishek and Norta, Alex and Draheim, Dirk},
  year    = {2021},
  journal = {ACM Computing Surveys},
  volume  = {54},
  number  = {5},
  pages   = {1--34}
}

@article{Tchakounte_2022_smart,
  title   = {A Smart Contract Logic to Reduce Hoax Propagation across Social Media},
  author  = {Tchakounté, Franklin and Amadou Calvin, Koudanbe and Ari, Ado Adamou Abba and Fotsa Mbogne, David Jaures},
  year    = {2022},
  journal = {Journal of King Saud University - Computer and Information Sciences},
  volume  = {34},
  number  = {6},
  pages   = {3070--3078}
}

@article{Ajienka_2020_empirical,
  title   = {An Empirical Analysis of Source Code Metrics and Smart Contract Resource Consumption},
  author  = {Ajienka, Nemitari and Vangorp, Peter and Capiluppi, Andrea},
  year    = {2020},
  journal = {Journal of Software: Evolution and Process},
  volume  = {32},
  number  = {10},
  pages   = {1--22}
}

@inproceedings{Ghaleb_2023_AChecker,
  title     = {{{AChecker}}: {{Statically Detecting Smart Contract Access Control Vulnerabilities}}},
  booktitle = {Proceedings of the {45th {IEEE}}/{{ACM}} {{International Conference}} on {{Software Engineering}} (ICSE)},
  author    = {Ghaleb, Asem and Rubin, Julia and Pattabiraman, Karthik},
  year      = {2023},
  pages     = {945--956}
}

@inproceedings{AlDebeyan_2022_Improving,
  title     = {Improving the Performance of Code Vulnerability Prediction Using Abstract Syntax Tree Information},
  booktitle = {Proceedings of the {{18th International Conference}} on {{Predictive Models}} and {{Data Analytics}} in {{Software Engineering}} (PROMISE)},
  author    = {Al Debeyan, Fahad and Hall, Tracy and Bowes, David},
  year      = {2022},
  pages     = {2--11}
}

@inproceedings{Wang_2022_Unified,
  title     = {Unified Abstract Syntax Tree Representation Learning for Cross-Language Program Classification},
  booktitle = {Proceedings of the {{30th IEEE}}/{{ACM International Conference}} on {{Program Comprehension}} (ICPC)},
  author    = {Wang, Kesu and Yan, Meng and Zhang, He and Hu, Haibo},
  year      = {2022},
  pages     = {390--400}
}

@article{Asudani_2023_Impact,
  title   = {Impact of Word Embedding Models on Text Analytics in Deep Learning Environment: A Review},
  author  = {Asudani, Deepak Suresh and Nagwani, Naresh Kumar and Singh, Pradeep},
  year    = {2023},
  journal = {Artificial Intelligence Review},
  volume  = {56},
  number  = {9},
  pages   = {10345--10425}
}

@article{Church_2017_Word2Vec,
  title   = {{{Word2Vec}}},
  author  = {Church, Kenneth Ward},
  year    = {2017},
  journal = {Natural Language Engineering},
  volume  = {23},
  number  = {1},
  pages   = {155--162}
}

@inproceedings{Yu_2021_Knowledge,
  title     = {Knowledge {{Embedding Based Graph Convolutional Network}}},
  booktitle = {Proceedings of the {{Web Conference}} (WWW)},
  author    = {Yu, Donghan and Yang, Yiming and Zhang, Ruohong and Wu, Yuexin},
  year      = {2021},
  pages     = {1619--1628}
}

@inproceedings{ibba2021evaluating,
  title     = {Evaluating machine-learning techniques for detecting smart ponzi schemes},
  author    = {Ibba, Giacomo and Pierro, Giuseppe Antonio and Di Francesco, Marco},
  booktitle = {Proceedings of the IEEE/ACM 4th International Workshop on Emerging Trends in Software Engineering for Blockchain (WETSEB)},
  pages     = {34--40},
  year      = {2021}
}

@article{kim2019scanat,
  title   = {ScanAT: identification of bytecode-only smart contracts with multiple attribute tags},
  author  = {Kim, Yuntae and Pak, Dohyun and Lee, Jonghyup},
  journal = {IEEE Access},
  volume  = {7},
  pages   = {98669--98683},
  year    = {2019}
}

@article{tann2018towards,
  title   = {Towards safer smart contracts: A sequence learning approach to detecting security threats},
  author  = {Tann, Wesley Joon-Wie and Han, Xing Jie and Gupta, Sourav Sen and Ong, Yew-Soon},
  journal = {arXiv preprint arXiv:1811.06632},
  year    = {2018}
}

@inproceedings{nikolic2018finding,
  title     = {Finding the greedy, prodigal, and suicidal contracts at scale},
  author    = {Nikoli{\'c}, Ivica and Kolluri, Aashish and Sergey, Ilya and Saxena, Prateek and Hobor, Aquinas},
  booktitle = {Proceedings of the 34th annual computer security applications conference},
  pages     = {653--663},
  year      = {2018}
}

@inproceedings{dhaya2014detecting,
  title     = {Detecting software vulnerabilities in android using static analysis},
  author    = {Dhaya, R and Poongodi, M},
  booktitle = {Proceedings of the IEEE International Conference on Advanced Communications, Control and Computing Technologies},
  pages     = {915--918},
  year      = {2014}
}

@inproceedings{liu2024efficient,
  title     = {An Efficient Cross-Contract Vulnerability Detection Model Integrating Machine Learning and Fuzz Testing},
  author    = {Liu, Huipeng and Cui, Baojiang and Xu, Jie and Niu, Lihua},
  booktitle = {Proceedings of the International Conference on Emerging Internet, Data \& Web Technologies},
  pages     = {297--306},
  year      = {2024}
}

@misc{etherscan,
  author       = {Tan Matai, Wei Chuan},
  title        = {Blockchain Explorer},
  howpublished = {Available: \url{https://cn.etherscan.io/}}
}

@inproceedings{feist2019slither,
  title     = {Slither: a static analysis framework for smart contracts},
  author    = {Feist, Josselin and Grieco, Gustavo and Groce, Alex},
  booktitle = {2019 IEEE/ACM 2nd International Workshop on Emerging Trends in Software Engineering for Blockchain (WETSEB)},
  pages     = {8--15},
  year      = {2019}
}

@inproceedings{li2024cobra,
  title     = {COBRA: Interaction-Aware Bytecode-Level Vulnerability Detector for Smart Contracts},
  author    = {Li, Wenkai and Li, Xiaoqi and Li, Zongwei and Zhang, Yuqing},
  booktitle = {Proceedings of the 39th IEEE/ACM International Conference on Automated Software Engineering (ASE)},
  pages     = {1358--1369},
  year      = {2024}
}

@inproceedings{kong2024characterizing,
  title     = {Characterizing the Solana NFT Ecosystem},
  author    = {Kong, Dechao and Li, Xiaoqi and Li, Wenkai},
  booktitle = {Companion Proceedings of the ACM on Web Conference (WWW)},
  pages     = {766--769},
  year      = {2024}
}

@inproceedings{niu2024unveiling,
  title     = {Unveiling Wash Trading in Popular NFT Markets},
  author    = {Niu, Yuanzheng and Li, Xiaoqi and Peng, Hongli and Li, Wenkai},
  booktitle = {Companion Proceedings of the ACM on Web Conference (WWW)},
  pages     = {730--733},
  year      = {2024}
}

@article{liu2025sok,
  title   = {SoK: Security Analysis of Blockchain-based Cryptocurrency},
  author  = {Liu, Zekai and Li, Xiaoqi},
  journal = {arXiv preprint arXiv:2503.22156},
  year    = {2025}
}

@article{bu2025smartbugbert,
  title   = {SmartBugBert: BERT-Enhanced Vulnerability Detection for Smart Contract Bytecode},
  author  = {Bu, Jiuyang and Li, Wenkai and Li, Zongwei and Zhang, Zeng and Li, Xiaoqi},
  journal = {arXiv preprint arXiv:2504.05002},
  year    = {2025}
}

@article{li2023overview,
  title   = {An overview of AI and blockchain integration for privacy-preserving},
  author  = {Li, Zongwei and Kong, Dechao and Niu, Yuanzheng and Peng, Hongli and Li, Xiaoqi and Li, Wenkai},
  journal = {arXiv preprint arXiv:2305.03928},
  year    = {2023}
}

@inproceedings{li2024stateguard,
  title     = {StateGuard: Detecting State Derailment Defects in Decentralized Exchange Smart Contract},
  author    = {Li, Zongwei and Li, Wenkai and Li, Xiaoqi and Zhang, Yuqing},
  booktitle = {Companion Proceedings of the ACM on Web Conference (WWW)},
  pages     = {810--813},
  year      = {2024}
}

@inproceedings{li2024defitail,
  title     = {DeFiTail: DeFi Protocol Inspection through Cross-Contract Execution Analysis},
  author    = {Li, Wenkai and Li, Xiaoqi and Zhang, Yuqing and Li, Zongwei},
  booktitle = {Companion Proceedings of the ACM on Web Conference (WWW)},
  pages     = {786--789},
  year      = {2024}
}

@article{mao2024scla,
  title   = {SCLA: Automated Smart Contract Summarization via LLMs and Control Flow Prompt},
  author  = {Mao, Yingjie and Li, Xiaoqi and Li, Wenkai and Wang, Xin and Xie, Lei},
  journal = {arXiv preprint arXiv:2402.04863},
  year    = {2024}
}

@inproceedings{li2020characterizing,
  title     = {Characterizing erasable accounts in ethereum},
  author    = {Li, Xiaoqi and Chen, Ting and Luo, Xiapu and Yu, Jiangshan},
  booktitle = {Proceedings of the 23rd International Conference on Information Security (ISC)},
  pages     = {352--371},
  year      = {2020}
}

@article{li2024guardians,
  title   = {Guardians of the ledger: Protecting decentralized exchanges from state derailment defects},
  author  = {Li, Zongwei and Li, Wenkai and Li, Xiaoqi and Zhang, Yuqing},
  journal = {IEEE Transactions on Reliability},
  year    = {2024}
}

@misc{li2021hybrid,
  author    = {Li, Xiaoqi},
  title     = {Hybrid analysis of smart contracts and malicious behaviors in ethereum},
  booktitle = {Hong Kong Polytechnic University},
  year      = {2021}
}

@misc{li2017discovering,
  author    = {Li, Xiaoqi and Yu, Le and Luo, Xiapu},
  title     = {On Discovering Vulnerabilities in Android Applications},
  booktitle = {Mobile Security and Privacy},
  year      = {2017},
  pages     = {155--166}
}

@inproceedings{10707457,
  author    = {Liu, Zekai and Li, Xiaoqi and Peng, Hongli and Li, Wenkai},
  booktitle = {2024 IEEE International Conference on Web Services (ICWS)},
  title     = {GasTrace: Detecting Sandwich Attack Malicious Accounts in Ethereum},
  year      = {2024},
  volume    = {},
  number    = {},
  pages     = {1409-1411}
}

@misc{wang2024smart,
  title         = {Smart Contracts in the Real World: A Statistical Exploration of External Data Dependencies},
  author        = {Wang, Yishun and Li, Xiaoqi and Ye, Shipeng and Xie, Lei and Ju Xing},
  year          = {2024},
  archiveprefix = {arXiv},
  eprint        = {2406.13253},
  journal       = {arXiv preprint}
}

@misc{li2025scalm,
  title         = {SCALM: Detecting Bad Practices in Smart Contracts Through LLMs},
  author        = {Li, Zongwei and Li, Xiaoqi and Li, Wenkai and others},
  year          = {2025},
  archiveprefix = {arXiv},
  eprint        = {2502.04347},
  journal       = {arXiv preprint}
}

@inproceedings{li2024detecting,
  author    = {Li, Wenkai and Liu, Zheng and Li, Xiaoqi and others},
  title     = {Detecting Malicious Accounts in Web3 through Transaction Graph},
  booktitle = {Proceedings of the 39th IEEE/ACM International Conference on Automated Software Engineering (ASE)},
  year      = {2024},
  pages     = {2482--2483}
}

@article{bu2025enhancing,
  title   = {Enhancing Smart Contract Vulnerability Detection in DApps Leveraging Fine-Tuned LLM},
  author  = {Bu, Jiuyang and Li, Wenkai and Li, Zongwei and Zhang, Zeng and Li, Xiaoqi},
  journal = {arXiv preprint arXiv:2504.05006},
  year    = {2025}
}

@inproceedings{wu2025atomicity,
  title   = {On the Atomicity and Efficiency of Blockchain Payment Channels},
  author  = {Wu, Di and Ren, Shoupeng and Bai, Yuman and He, Lipeng and Liu, Jian and Wen, Wu and Ren, Kui and Chen, Chun},
  journal = {Proceedings of the 34th USENIX Security Symposium (USENIX Security 25)},
  pages   = {4053--4072},
  year    = {2025}
}

@inproceedings{xiao2025parallelizing,
  title     = {Parallelizing Universal Atomic Swaps for $\{$Multi-Chain$\}$ Cryptocurrency Exchanges},
  author    = {Xiao, Danlei and Zhang, Chuan and Deng, Haotian and Liang, Jinwen and Wang, Licheng and Zhu, Liehuang},
  booktitle = {Proceedings of the 34th USENIX Security Symposium (USENIX Security 25)},
  pages     = {4073--4092},
  year      = {2025}
}

@inproceedings{aguilarSmartContractFamilies2024,
  title     = {Smart {{Contract Families}} in {{Solidity}}},
  booktitle = {Proceedings of the 34th {{International Conference}} on {{Collaborative Advances}} in {{Software}} and {{COmputiNg}} (CASCON)},
  author    = {Aguilar, Julio and Bak, Kacper and Boyle, Michael and Callens, Valerian and Gorzny, Jan},
  year      = 2024,
  pages     = {1--5}
}

@inproceedings{arceriSoundConstructionEVM2024,
  title     = {Towards a {{Sound Construction}} of {{EVM Bytecode Control-Flow Graphs}}},
  booktitle = {Proceedings of the 26th {{ACM International Workshop}} on {{Formal Techniques}} for {{Java-like Programs}} (FTfJP)},
  author    = {Arceri, Vincenzo and Merenda, Saverio Mattia and Dolcetti, Greta and Negrini, Luca and Olivieri, Luca and Zaffanella, Enea},
  year      = 2024,
  pages     = {11--16}
}

@inproceedings{caiEnablingCompleteAtomicity2024,
  title     = {Enabling {{Complete Atomicity}} for {{Cross-Chain Applications Through Layered State Commitments}}},
  booktitle = {Proceedings of the 43rd {{International Symposium}} on {{Reliable Distributed Systems}} ({{SRDS}})},
  author    = {Cai, Yuandi and Cheng, Ru and Zhou, Yifan and Zhang, Shijie and Xiao, Jiang and Jin, Hai},
  year      = 2024,
  pages     = {248--259}
}

@article{wang2024ContractsentryStaticAnalysis,
  title   = {Contractsentry: A Static Analysis Tool for Smart Contract Vulnerability Detection},
  author  = {Wang, Shiji and Zhao, Xiangfu},
  year    = 2024,
  journal = {Automated Software Engineering},
  volume  = {32},
  number  = {1},
  pages   = {1}
}

@article{grossmanPracticalVerificationSmart2024,
  title   = {Practical {{Verification}} of {{Smart Contracts}} Using {{Memory Splitting}}},
  author  = {Grossman, Shelly and Toman, John and Bakst, Alexander and Arora, Sameer and Sagiv, Mooly and Nandi, Chandrakana},
  year    = 2024,
  journal = {Artifact for our paper titled "Practical Verification Of Smart Contracts using Memory Splitting"},
  volume  = {8},
  number  = {OOPSLA2},
  pages   = {356:2402--356:2433}
}

@article{hanOSwapPreservingAtomicity2026,
  title   = {{{OSwap}}: {{Preserving}} the {{Atomicity}} and {{Indistinguishability}} of \textbackslash bm N\_\textbackslash bm 1\textbackslash bm \textbackslash sim \textbackslash bm N\_\textbackslash bm 2n1{$\sim$}n2 {{Swap Without Synchronous Blockchain Communication}}},
  author  = {Han, Panpan and Yan, Zheng and Yang, Laurence T. and Bertino, Elisa},
  year    = 2026,
  journal = {IEEE Transactions on Dependable and Secure Computing},
  volume  = {23},
  number  = {1},
  pages   = {477--490}
}

@article{jiaoSurveyEthereumSmart2024,
  title   = {A {{Survey}} of {{Ethereum Smart Contract Security}}: {{Attacks}} and {{Detection}}},
  author  = {Jiao, Tengyun and Xu, Zhiyu and Qi, Minfeng and Wen, Sheng and Xiang, Yang and Nan, Gary},
  year    = 2024,
  journal = {Distributed Ledger Technologies: Research and Practice},
  volume  = {3},
  number  = {3},
  pages   = {23:1--23:28}
}

@inproceedings{kumarVulnerabilitiesSmartContracts2024,
  title     = {``{{Vulnerabilities}} in {{Smart Contracts}}: {{A Detailed Survey}} of {{Detection}} and {{Mitigation Methodologies}}''},
  booktitle = {Proceedings of the 2024 {{International Conference}} on {{Emerging Technologies}} in {{Computer Science}} for {{Interdisciplinary Applications}} ({{ICETCS}})},
  author    = {Kumar, Nayantara K and Honnungar, Niranjan V and Sharwari Prakash, M and Lohith, J J},
  year      = 2024,
  pages     = {1--7}
}

@article{liASTRODetectingAccess2025,
  title   = {{{ASTRO}}: {{Detecting Access Control Vulnerabilities}} in {{Smart Contracts}} via {{Graph Similarity Comparison}}},
  author  = {Li, Wei and Nan, Yuhong and Ye, Mingxi and Zhang, Jingwen and Zheng, Peilin and Zheng, Zibin},
  year    = 2025,
  journal = {IEEE Transactions on Software Engineering},
  volume  = {51},
  number  = {12},
  pages   = {3267--3283}
}

@article{wangContractCheckCheckingEthereum2024,
  title   = {{{ContractCheck}}: {{Checking Ethereum Smart Contracts}} in {{Fine-Grained Level}}},
  author  = {Wang, Xite and Tian, Senping and Cui, Wei},
  year    = 2024,
  journal = {IEEE Transactions on Software Engineering},
  volume  = {50},
  number  = {7},
  pages   = {1789--1806}
}

@article{wangEmpiricalAnalysisSmart2026,
  title   = {Empirical {{Analysis}} of {{Smart Contract Factories}} on {{EVM-compatible Chains}}},
  author  = {Wang, Ziyue and Shen, Zongwen and Chen, Lei and Song, Wei and Ge, Jidong and Huang, LiGuo and Luo, Bin},
  year    = 2026,
  journal = {ACM Transactions on Software Engineering and Methodology}
}

@article{weiSurveyQualityAssurance2024,
  title   = {Survey on {{Quality Assurance}} of {{Smart Contracts}}},
  author  = {Wei, Zhiyuan and Sun, Jing and Zhang, Zijian and Zhang, Xianhao and Yang, Xiaoxuan and Zhu, Liehuang},
  year    = 2024,
  journal = {ACM Computing Surveys},
  volume  = {57},
  number  = {2},
  pages   = {32:1--32:36}
}

@article{zhuSurveySecurityAnalysis2024,
  title   = {A {{Survey}} on {{Security Analysis Methods}} of {{Smart Contracts}}},
  author  = {Zhu, Huijuan and Yang, Lei and Wang, Liangmin and Sheng, Victor S.},
  year    = 2024,
  journal = {IEEE Transactions on Services Computing},
  volume  = {17},
  number  = {6},
  pages   = {4522--4539}
}

@inproceedings{boi2024VulnHuntGPTSmartContract,
  title     = {{{VulnHunt-GPT}}: A {{Smart Contract}} Vulnerabilities Detector Based on {{OpenAI chatGPT}}},
  booktitle = {Proceedings of the 39th {{ACM}}/{{SIGAPP Symposium}} on {{Applied Computing}}},
  author    = {Boi, Biagio and Esposito, Christian and Lee, Sokjoon},
  year      = 2024,
  pages     = {1517--1524}
}

@inproceedings{hu2023LargeLanguageModelPowered,
  title     = {Large {{Language Model-Powered Smart Contract Vulnerability Detection}}: {{New Perspectives}}},
  booktitle = {2023 5th {{IEEE International Conference}} on {{Trust}}, {{Privacy}} and {{Security}} in {{Intelligent Systems}} and {{Applications}} ({{TPS-ISA}})},
  author    = {Hu, Sihao and Huang, Tiansheng and {\.I}lhan, Fatih and Tekin, Selim Furkan and Liu, Ling},
  year      = 2023,
  pages     = {297--306}
}

@article{wei2025AdvancedSmartContract,
  title   = {Advanced {{Smart Contract Vulnerability Detection}} via {{LLM-Powered Multi-Agent Systems}}},
  author  = {Wei, Zhiyuan and Sun, Jing and Sun, Yuqiang and Liu, Ye and Wu, Daoyuan and Zhang, Zijian and Zhang, Xianhao and Li, Meng and Liu, Yang and Li, Chunmiao and Wan, Mingchao and Dong, Jin and Zhu, Liehuang},
  year    = 2025,
  journal = {IEEE Transactions on Software Engineering},
  volume  = {51},
  number  = {10},
  pages   = {2830--2846}
}

@inproceedings{zhuang2021smart,
  title     = {Smart contract vulnerability detection using graph neural networks},
  author    = {Zhuang, Yuan and Liu, Zhenguang and Qian, Peng and Liu, Qi and Wang, Xiang and He, Qinming},
  booktitle = {Proceedings of the twenty-ninth international conference on international joint conferences on artificial intelligence (IJCAI)},
  pages     = {3283--3290},
  year      = {2021}
}

@article{10.1145/3702973,
  author    = {Chen, Chong and Su, Jianzhong and Chen, Jiachi and Wang, Yanlin and Bi, Tingting and Yu, Jianxing and Wang, Yanli and Lin, Xingwei and Chen, Ting and Zheng, Zibin},
  booktitle = {When ChatGPT Meets Smart Contract Vulnerability Detection: How Far Are We?},
  year      = {2025},
  volume    = {34},
  number    = {4},
  journal   = {ACM Transactions on Software Engineering and Methodology (TOSEM)},
  articleno = {100},
  pages     = {1--30}
}

@article{he2025VulnerabilityDetectionMethod,
  title   = {A {{Vulnerability Detection Method}} for {{Smart Contracts Based}} on {{Dynamic Meta Optimizer}}},
  author  = {He, Daojing and Gong, Wei and Chan, Sammy},
  year    = 2025,
  journal = {IEEE Internet of Things Journal},
  volume  = {12},
  number  = {13},
  pages   = {24904--24915}
}

@inproceedings{sikder2025EfficientAdaptationLarge,
  title     = {Efficient {{Adaptation}} of {{Large Language Models}} for {{Smart Contract Vulnerability Detection}}},
  booktitle = {Proceedings of the 21st {{International Conference}} on {{Predictive Models}} and {{Data Analytics}} in {{Software Engineering}}},
  author    = {Sikder, Fadul and Lei, Yu and Ji, Yuede},
  year      = 2025,
  pages     = {65--74}
}

@article{sun2025MTVHunterSmartContracts,
  title   = {{{MTVHunter}}: {{Smart Contracts Vulnerability Detection Based}} on {{Multi-Teacher Knowledge Translation}}},
  author  = {Sun, Guokai and Zhuang, Yuan and Zhang, Shuo and Feng, Xiaoyu and Liu, Zhenguang and Zhang, Liguo},
  year    = 2025,
  journal = {Proceedings of the AAAI Conference on Artificial Intelligence},
  volume  = {39},
  number  = {14},
  pages   = {15169--15176}
}

@article{yu2025SmartLLaMADPOReinforcedLarge,
  title   = {Smart-LLaMA-DPO: Reinforced Large Language Model for Explainable Smart Contract Vulnerability Detection},
  author  = {Yu, Lei and Huang, Zhirong and Yuan, Hang and Cheng, Shiqi and Yang, Li and Zhang, Fengjun and Shen, Chenjie and Ma, Jiajia and Zhang, Jingyuan and Lu, Junyi and others},
  journal = {Proceedings of the ACM on Software Engineering},
  volume  = {2},
  number  = {ISSTA},
  pages   = {182--205},
  year    = {2025}
}

@article{zenggang2025NDLSCNewDeep,
  title   = {{{NDLSC}}: {{A New Deep Learning-based Approach}} to {{Smart Contract Vulnerability Detection}}},
  author  = {Zenggang, Xiong and Qiangqiang, Lou and Youfeng, Li and Hao, Chen and Xuemin, Zhang and Yuan, Li and Jing, Li},
  year    = 2025,
  journal = {Journal of Signal Processing Systems},
  volume  = {97},
  number  = {1},
  pages   = {49--68}
}

@inproceedings{tas2024AtomicFairData,
  title     = {Atomic and {{Fair Data Exchange}} via {{Blockchain}}},
  booktitle = {Proceedings of the 2024 on {{ACM SIGSAC Conference}} on {{Computer}} and {{Communications Security}}},
  author    = {Tas, Ertem Nusret and Seres, Istv{\'a}n Andr{\'a}s and Zhang, Yinuo and Melczer, M{\'a}rk and Kelkar, Mahimna and Bonneau, Joseph and Nikolaenko, Valeria},
  year      = 2024,
  pages     = {3227--3241}
}

\end{document}